# The Benefits and Harms of Transmitting Into Space

Jacob Haqq-Misra[a,*], Michael W. Busch[a,b], Sanjoy M. Som[a,1], and Seth D, Baum[a,c]

    a. Blue Marble Space Institute of Science
    b. Department of Earth and Space Sciences, University of California Los Angeles
    c. Department of Geography, Pennsylvania State University
    * Corresponding author. Email: jacob@bmsis.org
    1. Present address: Exobiology Branch, NASA Ames Research Center

**Abstract**
Deliberate and unintentional radio transmissions from Earth propagate into space. These transmissions could be detected by extraterrestrial watchers over interstellar distances. Here, we analyze the harms and benefits of deliberate and unintentional transmissions relevant to Earth and humanity. Comparing the magnitude of deliberate radio broadcasts intended for messaging to extraterrestrial intelligence (METI) with the background radio spectrum of Earth, we find that METI attempts to date have much lower detectability than emissions from current radio communication technologies on Earth. METI broadcasts are usually transient and several orders of magnitude less powerful than other terrestrial sources such as astronomical and military radars, which provide the strongest detectable signals. The benefits of radio communication on Earth likely outweigh the potential harms of detection by extraterrestrial watchers; however, the uncertainty regarding the outcome of contact with extraterrestrial beings creates difficulty in assessing whether or not to engage in long-term and large-scale METI.

## 1. Introduction

Does transmitting radio messages into space pose a risk to human civilization? Efforts to send messages to potential extraterrestrial watchers[1] have raised concerns that such actions may provoke unwanted attention. Similar transmissions into space, though unintentional, occur as a result of radio communication on Earth, and pose similar risks. This paper analyzes deliberate and unintentional transmissions into space and the degree to which these activities could provide benefits or harms to Earth and humanity.

    Electromagnetic waves have been used to communicate for over one hundred years. Television broadcasts, mobile phone conversations, satellite transmissions, and military, civil, and astronomical radars all use some part of the electromagnetic spectrum—particularly radio and microwave wavelengths—to transmit encoded information from a sender to a watcher. These technologies have transformed communication across the globe and have enabled human space flight and robotic exploration of the solar system. Nearly all terrestrial electromagnetic transmissions used for communication also radiate into space. Although such signals decrease in intensity as they move away from Earth, this *leakage radiation* can be detected over interstellar distances with a sufficiently sensitive telescope [1,2].

    Cocconi and Morrison [3] first suggested that a search for interstellar radio transmissions could uncover evidence of intelligent extraterrestrial life elsewhere in the galaxy. Over fifty years later, the search for extraterrestrial intelligence (SETI) has found no evidence of artificial signals in space, although efforts to broaden the search continue [4]. Another way to search for intelligence elsewhere in the universe involves transmitting messages toward target star systems.

---

[1] Throughout this paper we use the term *watcher* to designate the recipient of an electromagnetic signal, although the term *observer* can be used interchangeably.

This is known as "messaging to extraterrestrial intelligence" or METI [5]. The ultimate goal of METI is to transmit a signal that is eventually received by an extraterrestrial civilization, although the vast distances between stars renders any conversation a multigenerational project [6]. Nevertheless, a handful of attempts at METI have been made over the last half century with messages increasing in size and complexity [7]. These efforts can be considered as symbolic or demonstrations of human technology rather than serious efforts to converse with extraterrestrial civilizations.

Both deliberate METI signals and unintentional leakage radiation contribute to the overall radio[2] emission from Earth. There has been concern that this signature of our technological civilization could constitute a risk because it reveals our location in the galaxy to any potentially hostile extraterrestrial civilizations [8-18]. There have even been calls for a moratorium on deliberate METI transmissions until international agreements for how to proceed have been reached [19]. Others have argued that METI broadcasts do not pose a significant risk [7,20-23] because any extraterrestrial watchers would be able to establish the presence of life on Earth by the spectrum of reflected ultraviolet, optical, and near-infrared sunlight into space from the surface and the atmosphere. An extraterrestrial watcher could also potentially learn of our technological civilization by detecting artificial nighttime lighting of large urban areas [24].

Optimists suggest that contact with extraterrestrials could bring about great benefits for humanity [25], while others note that contact with technological civilizations has often resulted in the collapse of stone-age societies on Earth [14]. Contact with extraterrestrials could result in a number of consequences [26], so if the risk from transmission into space is non-zero, then should transmissions into space be permitted, regulated, or banned? If human activities can be detected across astronomical distances, then should humanity cease or attempt to disguise such actions? Does METI significantly increase risks to Earth and human civilization? These questions have been raised repeatedly in the research literature as well as in media and political coverage of SETI and METI research. We address these questions in this paper by reviewing existing knowledge of the Earth's radio signature, which includes the relative strength of signals potentially detectable over interstellar distances. We then develop an analytical framework for evaluating the consequences of transmission and discuss this analysis in the context of existing policies and protocols.

## 2. Detectability of Radio Transmissions from Earth

Before about one hundred years ago, Earth was "radio quiet" with no significant emission of radio waves compared to other objects in the Solar System (particularly the Sun and the gas giant planets). The development of radio transmitters initiated a new era where the technological activity of humans altered the electromagnetic spectrum of Earth. Other changes in Earth's spectrum driven by its biosphere include the rise in atmospheric oxygen about 2.4 billion years ago [27] and the proliferation of photosynthesis [28]. However, these changes to the spectrum primarily occurred in the near-infrared to ultraviolet regions of the electromagnetic spectrum where the planet is brightest. By contrast, in the radio and microwave regions of the electromagnetic spectrum the Earth was previously very faint.

---

[2] Hereafter, we will use the term *radio* to describe electromagnetic radiation at frequencies greater than 10-30 MHz (which is the cutoff frequency for radiation to penetrate the ionosphere) and less than ~100 GHz (where atmospheric absorption becomes prohibitively high). This range of frequencies includes *microwaves* as well as lower radio frequencies.

Earth's radio leakage comes from many different sources that range from active cell phones to television and radio broadcasts to high-power radars used for astronomy and by the military.[3] All of these signals travel through space at the speed of light, so television broadcasts that occurred twenty years ago are now twenty light years away from Earth (for comparison, Proxima Centauri, the closest star to the Sun, is 4.2 light years away). Leakage radiation from television transmitters occurs approximately in a sphere surrounding Earth, so that the distance at which Earth's radio signature can be detected has sometimes been termed the *radiosphere*. However, radar beams are the strongest radio leakage, and spread into space from Earth like pins on a pincushion, with most of the beams (pins) concentrated in the northern hemisphere. The intensity of signals from Earth decays with distance according to an inverse square law, but prior analyses have shown that these faint signals could still be detected at astronomical distances by a sensitive receiver and a sufficiently large antenna [1,2].

To determine if a given transmission can be detected at a given distance, some assumptions must be made about the receiving radio telescopes. To quantify the relative detectability of different types of leakage, we assume a watcher equipped with a radio telescope or radio telescope array with high angular and frequency resolution. This is because with low resolution in either angle or frequency, background galactic radio emission dominates the leakage radiation. This is quantified by comparing the spectral flux density (power per unit area per unit frequency) of the galactic background at a watcher's antenna with the flux of the leakage from Earth [29]. On the other hand, with high resolution only a very small fraction of the radio background overlies the leakage radiation and only the properties of the leakage radiation itself matter. In this case, a watcher will be able to detect and potentially interpret signals from Earth as long as the *number of photons* per unit area of antenna per bit of data is significantly greater than the unavoidable thermal noise in their receivers. Here we can express a single bit of data in terms of the bandwidth $B$ of the signal as a time equal to $1/B$. The thermal noise in the receiver's detectors in this time and bandwidth will be proportional to $B$, meaning that broadband signals—such as television transmissions, cellphone networks, and wireless Internet—are more difficult to detect.[4]

The relevant quantities for a transmitting antenna are the gain (effectively the fraction of the sky over which the antenna transmits), the transmitter power, and the choice of broadcast frequency. For a transmitting antenna with gain $G$ and power $P$ that operates at frequency $v$, the ratio of the signal to the receiver noise per unit area of the watcher's *receiving* antenna is proportional to $PG/Bvr^2$, where $r$ is the distance from the transmitter to the receiver. There is therefore a limiting distance $r_l$ for detectability of

$$r_l \propto P^{1/2} G^{1/2} B^{-1/2} v^{-1/2}. \tag{1}$$

Note that Eq. (1) is a proportionality, rather than an equality; the true value of the distance $r_l$ depends on the collecting area and signal to noise threshold of the receiving antenna. A signal transmitted from Earth traverses a cone, with the vertex at Earth. The volume of the cone, and the volume over which the signal will be detectable is $V \propto r_l^3/G$, or

$$V \propto P^{3/2} G^{1/2} B^{-3/2} \nu^{-3/2} .\qquad(2)$$

---

[3] Civilian air traffic radars used for local navigation are much less powerful and have much lower gains than radars used for astronomy or to track spacecraft and intercontinental ballistic missiles. The ranges required for, e.g., air traffic control civilian radar are hundreds of times smaller and require much less power, while the requirement to scan all of the the local airspace requires much lower gain. We will therefore neglect consideration of civilian navigation radar in our analysis of leakage radiation.

[4] We here refer to the *detectability* of a signal, rather than to the ability of a watcher to interpret it. An encrypted or compressed signal may be detectable but not intelligible. See Sec. 3 and Sec. 5 for further discussion.

For example, the Arecibo Planetary Radar typically transmits at a power of 0.8 MW and a frequency of 2380 MHz, with a gain of ~$10^8$ (see Table 1). This means that low bandwidth transmissions from Arecibo, with $B \sim 0.1$ Hz, would be detectable by a watcher with a 1 km² receiving antenna at distances up to 200,000 light years, while high bandwidth signals, with $B \sim 10^7$ Hz, would be detectable out to about 5 light years by the same watcher. By comparison, television carrier waves have similar power but gain ~10, $B \sim 1$ Hz, and frequencies in the range of 100 to 2700 MHz; such signals could be detected with a square kilometer array out to a distance of about 50 light years.

**Table 1: Relative Detectability of Sources of Radio Leakage**

| Transmitter | T | P(W) | G | B(Hz) | ν(MHz) | V(norm) | $r_1$ (ltyr) for 1 km² array |
|---|---|---|---|---|---|---|---|
| Broadband (Cell phones and TV)[a] | 1 | $10^8$ | 10 | ~$10^9$ | 100 – 2700 | <$10^{-11}$ | 0.03 |
| TV Carrier Wave (one station) | 1 | $3 \times 10^5$ | 10 | 1 | 100 – 2700 | $3 \times 10^{-2}$ | 50 |
| Arecibo Radar CW       Broadband | $10^{-2}$  " | $8 \times 10^5$  " | $10^8$  " | 0.1 $2 \times 10^7$ | 2380  " | 1 $4 \times 10^{-13}$ | 200,000 4.5 |
| USAF PAVE PAWS Military Radar | 0.1 | $10^5$ | $3 \times 10^4$ | 0.1 | 430 | $10^{-1}$ | 60,000 |
| Goldstone CW       Broadband | $10^{-2}$  " | $4 \times 10^5$  " | $10^8$  " | 0.1 $4 \times 10^7$ | 8650  " | $10^{-1}$ $10^{-14}$ | 90,000 4.5 |
| Evpatoria CW | $10^{-3}$ | $10^5$ | $3 \times 10^7$ | 0.1 | 6000 | $3 \times 10^{-4}$ | 20,000 |
| Lone Signal at Madley 49 | 0.1 | $10^4$ | $10^6$ | $6 \times 10^4$ | 6000 | $6 \times 10^{-13}$ | 1.5 |

Transmitter properties are approximate, but provide order-of-magnitude estimates of the relative detectability of different forms of radio leakage. The detectable volumes have been normalized to that of the Arecibo Planetary Radar, and the limiting radius for detectability is given in lightyears for a hypothetical watcher with a 1 km² receiving array that provides high spatial resolution. For the radars, the bandwidth of continuous wave (CW) transmissions is limited to ~0.1 Hz at GHz frequencies by the Earth's rotation.
*Sources*: Broadband leakage & TV carrier waves - ITU, Arecibo - www.naic.edu, PAVE PAWS fact sheet, Goldstone - NASA DSN, Evpatoria RT-70 radio telescope factsheet, pers. comm. from Lone Signal
[a] Because this leakage is broadband, this is an extreme upper bound (see text).

Most sources of leakage radiation are transient. The volume over which they are detectable, and often the time for which they are detectable at any one point in space, is directly proportional to how much time they transmit for. We can account for this by including an additional factor *T*:

$$V \propto T P^{3/2} G^{1/2} B^{-3/2} \nu^{-3/2} .\qquad(3)$$

*T* is the fraction of time that the transmitter is on so that *T = 1* implies continuous transmission. For example, Arecibo transmits about 2% of the time in a year. If it transmitted 4% of the time, it would cover twice as many directions and hence twice as much volume. (This assumes that the transmission directions are uncorrelated with each other.) In the case of the TV carrier waves, which are always on and sweep the sky every day, the total detectable volume is approximately spherical and the detectable time for such a signal at any given point is proportional to the absolute amount of time that the transmitter operates. Our scaling relationships are approximate and most applicable in the limit of high gain, where each transmission is a narrow cone. We use this approximation because the most detectable leakage from Earth is from high-gain transmitters.

A watcher could potentially detect a signal with an apparent brightness of less than one photon per bit by watching for a longer period of time and measuring the average transmitted power. Such a weak signal *cannot* be interpreted because there is insufficient information to establish the value of each bit. The relative detectability of the different sources of leakage at such large distances remains as in Eq. (3) because the receiver noise overlying the signal still depends on its bandwidth, although the absolute value of $V$ is higher in this case.

The chance of a signal actually reaching an extraterrestrial watcher is related to the number of stars within the detectable volume, assuming that any watchers are on or nearby extrasolar planets. This is proportional to $V$ as long as $r_l$ < 500 light years. At greater distances, the thickness of the galactic plane becomes significant and the number of stars inside the detectable volume depends on the transmission direction of the antenna. The habitability of stellar systems may also vary as a function of location in the Galaxy [30], so transmissions preferentially directed away from the galactic plane (such as those from Arecibo) may be less likely to be detected. However, many sources of leakage do not cluster significantly on or away from the galactic plane, so the number of stars remains roughly proportional to $V$ itself.

Using Eq. (3), we can quantify the relative detectability of different sources of leakage (Table 1). Since the seminal study of radio leakage by Sullivan et al. in 1978 [1], broadband radio leakage has dramatically increased due to the introduction of cell phones, digital television, and extensive satellite communication traffic. However, these signals are detectable over only a very small volume because of their large bandwidths. The galactic radio background also cannot be neglected for broadband signals, so the values given in Table 1 are upper bounds. The carrier waves of television broadcasts are far more detectable, even as broadcasters are switching from analog to digital modulation of their transmitters, which leads to a decrease in the fraction of transmitted power by television carrier waves [31]. As an illustration, amateur radio observers are able to track television carrier waves reflected off of the Moon back towards Earth [32].

As noted by Sullivan et al. [1], television carrier waves are significantly patterned. Most transmitters are over land and direct their transmissions primarily towards the local horizon. Because each point on the sky crosses the horizon twice per day (once when it rises and once when it sets), each transmitter sweeps across every point on the sky that is visible at its latitude twice per day (aside from transmitters at high latitudes where certain points never set below the horizon). The detectable volume of the carrier waves covers all of space near Earth, and a patient watcher could map out the transmitters' positions in longitude. Recent SETI messaging campaigns, such as the Cosmic Call initiative at Evpatoria [5] and the Lone Signal project using the Madley 49 antenna,[5] have much higher gain than the carrier waves, but also much lower transmitting power, and relatively short transmission duration, so they are much less detectable. The most detectable sources of radio leakage on Earth are radar transmissions from Arecibo, the Goldstone Deep Space Network site, and various military installations. They have high gain, high power, and often transmit very low bandwidths, so they are detectable at very great distances over a small set of directions. However, television carrier waves could be considered as equally important because they cover all of the space nearby Earth.

In addition to the total volume, duration is an important factor for detectability. A message transmitted at just one moment is much less likely to be received than a message transmitted continuously. All previous METI attempts transmitted for a short time and are thus unlikely to be received compared with the radio leakage, which has persisted continuously for several decades (and counting). Effective METI strategy requires long-term commitment to transmitting for at least hundreds or thousands of years, depending on the distance to the nearest watchers [6,33]. Furthermore, individual METI attempts occur at a particular frequency or set of frequencies, which in turn requires that any watcher must be sensitive to that frequency. Many suggestions have been made as to the most likely frequencies to be used for interstellar communication [5,6,33,29,34]. Radio leakage, on the other hand, occurs at a wide range of frequencies, which may increase the likelihood of these signals being noticed. It is much more likely that any watchers will be able to, or have, detected the radio leakage than any METI transmissions to date.

---

[5]Publications about the Lone Signal METI project are currently unavailable. For more information, contact: Lone Signal LLC, 131 Sullivan Street #2B, New York, NY 10012.

## 3. The Possibility of Extraterrestrial Watchers

For extraterrestrial intelligence (ETI) to cause any harms or benefits associated with humanity transmitting into space, there must be ETI elsewhere in the universe, and particularly elsewhere within the Milky Way. Star systems with planets are plentiful in the galaxy, with current transit surveys discovering rocky planets close in size to, and even smaller than, Earth [35-37]. Many of these planets reside within the habitable zone of their parent stars, so that liquid water (or perhaps some other solvent) could accumulate on their surfaces, which may lead to the emergence of living systems and the eventual development of intelligent, technological life. This process began at least once in the galaxy, on Earth, roughly 4 billion years ago. However, even though life is potentially common in the galaxy, no evidence of extraterrestrial civilization has yet been discovered.

If life is widespread in the galaxy, and intelligence is a common consequence of life, then why has extraterrestrial intelligence not been detected? This question is historically known as the *Fermi paradox* (named after physicist Enrico Fermi) and speculates that a civilization with only modest advances beyond our own could easily colonize the entire galaxy in a fraction of its current age. The absence of such a civilization could imply that Earth is the only planet with intelligent life [38]. However, many other resolutions to the Fermi paradox have been proposed [39,40]. These include the possibilities that complex life is a rare phenomenon [41], intelligent societies tend toward self-destruction [6,8], extraterrestrials are hiding from us [42,43], or extraterrestrial civilizations grow too slowly to have reached Earth [44,45]. Because we have not yet observed any ETI, Fermi's question still remains unresolved. The idea that extraterrestrial civilizations could be watching for radio transmissions or broadcasting their own follows from our own use of radio technology. This does not imply that the evolution of life always produces a species capable of radio communication, but even the Fermi paradox does not preclude the possibility.

A standard thought experiment about the number of detectable civilizations in the galaxy was developed by Frank Drake, Carl Sagan, and others, at a 1961 meeting in Green Bank, West Virginia. The Drake Equation is a probabilistic expression to assess factors that contribute to the evolution of a detectable civilization and can be written as

$$N = R \cdot f_p \cdot n_e \cdot f_l \cdot f_i \cdot f_c \cdot L \tag{4}$$

Here $N$ is the present number of civilizations in the galaxy. It is the product of the rate of star formation in the galaxy $R$, the fraction of stars with planets $f_p$, the average number of potentially habitable planets in each planetary system $n_e$, the fraction of habitable planets that develop life $f_l$, the fraction of life-bearing planets that develop intelligent life $f_i$, the fraction of intelligence-bearing planets that develop and use detectable technology $f_c$, and the average lifetime of technological civilizations $L$.

Some of these terms are more easily estimated than others. Present observations suggest $R \approx 10$ stars per year and $f_p \geq 0.5$. Estimating the value of $n_e$ is more difficult because 'potentially habitable' can be defined in multiple ways [30,46-50], although the ongoing discovery of extrasolar planets will help to constrain this parameter. The remaining terms are the most difficult to estimate because life on Earth is the only known example. Humanity's ability to ponder life in the universe is contingent upon being part of an intelligent civilization. Failing to account for this contingency can lead to 'observation selection bias' in estimates of $f_l, f_i, f_c,$ and $L$ [17,51-53]. One of the more provocative terms is $L$, the lifetime of technological civilizations[6] since the advent of radio communication. If $L$ is low, then the resolution of the Fermi paradox is that detectable technological civilizations never exist long enough[7] to colonize space or send interstellar messages. The Drake equation provides a way to think about the factors involved in the search for extraterrestrial intelligence, but cannot tell us the actual value of $N$. At most, it gives a range of plausible values from 1 or less (Earth is unique in the galaxy and the nearest current ETI civilization is most likely millions of light years away in another galaxy) to ten thousand or more (ETI civilizations are abundant in the galaxy), depending on the values of the various parameters.

Instead of attempting to defend any particular estimate, we have tabulated a few different examples in Table 2, including the original Green Bank estimate of $N = 10$ and Drake's current estimate of $N = 10,000$. We have also included the *Rare Earth* position [41] that Earth-like planets and the evolution of complex life are phenomena and $N = 1$. This wide range illustrates our lack of knowledge about ETI: the galaxy could be teeming with ETI civilizations or humanity could be completely alone. If humans are in fact the only form of technological life in the galaxy, then there is zero chance of Earth transmissions being detected, at least within the next several million years ($N$ is the average number of civilizations in the *galaxy*, rather than the universe, so the nearest ETI are likely in the Andromeda galaxy if $N = 1$). However, if $N$ is large, then it is likely that there are ETI civilizations that could detect transmissions from Earth.[8]

---

[6] Strictly speaking, $L$ describes the length of time that a technological civilization emits detectable signals into space; however, this definition assumes that a technological civilization will continue to use radio indefinitely once it has been invented.

[7] One explanation for a low $L$ value is that technological civilizations inevitably destroy themselves. Alternatively, civilizations could cease to use detectable technology, producing an effectively lower $L$ even as the civilizations continue to exist.

[8] It is not necessarily the case that $N$ is always correlated with detectability. For example, perhaps there is only one ETI civilization in the Galaxy (so that $N = 2$), but this civilization has colonized a significant fraction of stellar systems. In this case, a single large civilization may be more likely to detect transmissions from Earth than several smaller civilizations that occupy a smaller fraction of stellar systems. However, given the absence of knowledge regarding ETI in the galaxy, we will assume in this paper that greater $N$ correlates with greater detectability.

**Table 2: Estimates of the Drake Equation**

|  | R (*/yr) | $f_p$ | $n_e$ | $f_l$ | $f_i$ | $f_c$ | L (yr) | N | $r_{nearest}$ (ltyr) |
|---|---|---|---|---|---|---|---|---|---|
| Green Bank (1961) | 10 | 0.5 | 2 | 1 | 0.01 | 0.01 | $10^4$ | 10 | ~15,000 |
| Rare Earth | 10 | 1 | 0.1 | 0.1 | 0.1 | 0.1 | $10^4$ | 1 | ~2,200,000 |
| Drake's Current Estimate | 10 | 1 | 1 | 0.3 | 0.3 | 1 | $10^4$ | $10^4$ | ~2500 |

Values of *N* for various possible combinations of parameters in the Drake equation. *rnearest* is the distance from us to the nearest alien civilization, assuming that civilizations are uniformly distributed across all stars.

ETI that exist at a location in time and space that permits detection of radio waves from Earth still face the task of actually detecting the signal. Intelligent, technological civilizations might miss a signal emanating from Earth because they decline to build or maintain sufficiently large radio telescopes, perhaps because they consider such endeavors unimportant [4,16,54] Even given a radio telescope or radio telescope array with high angular and spectral resolution and a large enough collecting area, when a signal arrives, the telescope must be pointed in the correct direction and be tuned to frequencies covering the bandwidth of the signal. These considerations are similar to the parameter $f_c$ in the Drake Equation (Eq. 4). We can therefore consider the ability of ETI to detect a signal by introducing an additional parameter: we let $f_d$ be the fraction of civilizations that maintain long-term broadband radio telescope facilities that would be capable of detecting signals of a particular brightness. This term $f_d$ depends on the nature of ETI across the galaxy. Because ETI have never been observed, the information to estimate this parameter is lacking.

For detection to actually occur, a fourth condition is necessary: the signal must be interpreted by ETI as having originated from another intelligent civilization. It is difficult to assess whether ETI watchers would be able to decipher or comprehend the transmitted information. Mathematical discourse has been suggested as the message content most likely to be understood [55-57] because mathematics concepts might be the most universally accessible. Intentionally transmitted signals built upon a computational language [58] also may have a higher likelihood of the information being understood by the receiver. By contrast, radio leakage of encrypted and/or complex multimedia messages or other culturally dependent messages will likely be meaningless to ETI because of the lack of a common framework for comprehension [7]. However, Earth's radio leakage and deliberate transmissions will likely be identifiable by ETI as a technological signature because no other examples of such signals exist in nature. The ability of ETI to decipher or interpret the content of a signal is therefore irrelevant to their ability to use it to learn that humans exist, so $f_d$ describes the fraction of civilizations that can detect any transmissions from Earth without regard to decipherability.

The probability of an Earth-originating signal being detected by ETI and affecting humanity by a response depends on both *N* and $f_d$. We can use these variables along with the detectable volume from Eq. (3) to express the probability that ETI detect a signal from Earth. Approximating civilizations as uniformly distributed we can write the probability $p_d$ that ETI detect a signal as

$$p_d = \frac{V}{V_g} N f_d , \quad (5)$$

where $V_g$ equals the total volume of the galaxy. Eq. (5) illustrates how the probability of ETI finding us depends on *N*, $f_d$, and *V*. The values of *N* and $f_d$ are unknown and are also beyond the influence of human civilization. In the absence of dependable estimates for $N$ and $f_d$, we cannot estimate $p_d$ with any accuracy, which makes any answers to questions about the value of transmitting signals from Earth speculative at best. Nevertheless, the only way for humanity to increase or decrease the probability of detection by ETI watchers is to change $V$ by increasing or decreasing the detectability of our radio emissions.

**4. The Value of Transmitting**
For detection of a signal by ETI to provide harms or benefits to humanity, the watchers must react in some way that would affect human society. The watching ETI civilization could do nothing, treat the detection exclusively as a scientific discovery, or something else that has no impact. These responses all result in zero value to Earth and humanity. ETI could also respond by transmitting back towards Earth, by sending exploratory probes or other objects here, or by visiting themselves. Each of these scenarios could bring benefits or harms to humanity.

A standard definition of risk is the probability of an event occurring multiplied by the magnitude of the harm from that event if it does occur. To estimate the risk of transmitting, we need estimates for the probability of detection by ETI (Eq. 5), the probability of response by ETI, and the magnitude of the response from ETI in terms of benefits or harms to humanity.

Assuming ETI have already discovered a signal and identified its source location, humans must recognize that contact with ETI might not be harmful. Humanity could benefit, or the effect could be neutral, as discussed in the scenario analyses of Baum et al. [26]. Let the term $p_B$ describe the probability of ETI response in a way that would benefit humanity, and let $p_H$ describe the probability of harmful ETI response. Likewise, let $p_N$ describe the probability of neutral ETI response that neither benefits nor harms humanity. Each of these probabilities ranges from 0 to 1. Beneficial, harmful, and neutral scenarios describe the set of all outcomes, so there is an additional constraint: $p_B + p_H + p_N = 1$.

Assuming that ETI exist, have detected a signal from Earth, and are responding in some way that would affect humanity, what kind of response could humanity expect? What would the magnitude of harms or benefits to humanity be? These questions are difficult to answer definitively because they depend on characteristics of ETI that we can only speculate about. We can only suggest possibilities and develop a framework for discussion.

Whether the encounter benefits or harms humanity will depend on the ETI's ethics. If interstellar travel is overly difficult or costly, then any interaction with ETI will be limited to exchanging messages. If communication is limited to remote messaging, then ETI probably will have little to gain in harming a human civilization that it cannot otherwise interact with. A non-malicious message would likely benefit humanity by providing information about ETI, although the discovery of ETI would also be a profound notion that would have positive and negative impacts for many societal structures, religions, and philosophies [13,59-61]. Perhaps, as Diamond [14] suggests, ETI would be as harmful to humanity as some groups of humans have been to other, technologically-inferior, groups of humans. Or perhaps, as Sagan and Newman [25] suggest, ETI would be beneficial to humanity, having outgrown any tendencies to act aggressively toward other civilizations. ETI could also have some other ethical disposition [62].

We can extend our framework to include the magnitude of ETI response to signal detection. Let $M_B$ describe the magnitude of beneficial ETI response (or the mean magnitude across a range of beneficial scenarios), where $M_B > 0$. Likewise, let $M_H$ describe the magnitude of harmful ETI response, where $M_H < 0$. The magnitude of neutral contact is $M_N = 0$ because there is no impact from detection. We can use these values for the magnitude of ETI responses to write an expression for the expected value $\langle M \rangle$ of the magnitude of ETI response:

$$\langle M \rangle = p_B M_B + p_N M_N + p_H M_H = p_B M_B + p_H M_H . \quad (6)$$

If the overall expected magnitude is beneficial to humanity then $\langle M \rangle > 0$, while if the overall expected magnitude is harmful to humanity then $\langle M \rangle < 0$. Because humanity has never observed ETI, the values of $p_B$, $p_H$, and $p_N$, and of $M_B$, $M_N$, and $M_H$ are unconstrainable. As such, we do not attempt to speculate on their magnitudes. Certain questions about the consequences of transmissions from Earth remain unanswerable.

Other factors involved in assessing the value of transmission into space include the benefits of conducting METI, the costs of conducting METI, and the value of radio communication on Earth. Even if no response is ever received, some METI projects can provide value to society by engaging the public through education and outreach programs. For example, the "Earth Speaks" website of the SETI Institute collects and analyzes messages from users around the world to identify major themes and ideas that people address [63], which may be useful data for psychologists or anthropologists Furthermore, the "Teen Age Message" transmitted from Evpatoria [5] includes content from a group of Russian teenagers, and the Lone Signal initiative contains messages from users worldwide—these projects not only increase the possibility of detection by ETI watchers but also engage the public in thinking about METI. Successful communication between cultures on Earth is arguably a first step toward successful message construction for METI, and the DearET project [7] aims to create a system for exchanging encoded messages across human cultural boundaries. This will identify potential content for future METI attempts while also engaging human cultures in conversation with one another. Technical research into SETI has led to advances in signal processing and distributed computing, such as the SETI@home project [64], that have impacted scientific research in many other disciplines; similar technological advances could arise if humanity someday chooses to engage in long-term and large-scale METI. No large-scale attempts at METI have endured for any significant length of time yet, so it is difficult to predict the technological or societal benefits of such a project prior to contact with ETI. Nevertheless, any serious attempt at METI will evoke interest and stimulate imagination in the public, at the very least. Considering factors such as these, we let $M_M$ describe the magnitude of benefits of conducting METI without regard to whether or not a response is received.

While some METI projects focused on education have relatively low costs, actual attempts at long-term transmitting will be costly to maintain for the multigenerational time periods required for communication [6,33]. We account for this using $M_C$ to describe the costs of conducting METI. Even without deliberate METI broadcasts, the leakage radiation could be detected by extraterrestrial watchers, so it is worth considering the relative value of these communication activities on Earth to the possible risk from detection by ETI. We describe the value of radio communications on Earth with the term $M_R$.

With these additional terms, we can write expressions for the value of METI and of radio communication. For METI, we can write the value $U_M$ as

$$U_M = p_d \langle M \rangle + M_M - M_C . \quad (7)$$

The total value of METI equals the consequences of ETI detection of a signal plus the other benefits of METI, minus the costs of METI. Likewise, for radio communication, we can write the value $U_R$ as

$$U_R = p_d \langle M \rangle + M_R . \quad (8)$$

The total value of radio communication equals the consequences of ETI detection of a signal plus the benefits of radio communication on Earth.

These expressions allow us to compare the relative value of METI with normal uses of radio. If the consequences of radio contact with ETI would bring about consequences much more costly than the present value of radio technology on Earth (that is, if $U_R << 0$), then ceasing all radio communication and becoming "radio quiet" would be a preferable course of action. However, if the benefits of radio communication are more valuable (that is, if $U_R >> 0$), then radio communication should continue on Earth today.

We can estimate the distance from us to the nearest extraterrestrial civilization for a particular set of Drake Equation parameters. If we assume that civilizations are uniformly distributed across all stars in the galaxy, then a given value of $N$ corresponds to an average distance $r_{nearest}$ between Earth and the nearest other technological civilization (Table 2). When $L$ (measured in years) is less than $r_{nearest}$ (measured in light years), any watchers will likely receive terrestrial messages after human extinction, so that SETI and METI are analogous to the exchange of time capsules rather than conversation. In the *Rare Earth* case where $N = 1$, Earth likely holds the only civilization in this galaxy, and the nearest extraterrestrial civilization at the present time is likely to be in the Andromeda Galaxy which, at a distance of 2.5 million light years, is the nearest large galaxy to the Milky Way. SETI and METI attempts are more likely to succeed if the value of $N$ is large in the Milky Way, but there can be no conversation with ETI unless $L$ is large.

If $L$ is small in the galaxy and human civilization will follow an average course of development, then it is unlikely that METI would provide any benefits or harms from contact with ETI. In other words, if civilizations are short-lived, then there will not be enough time for interstellar communication, let alone interstellar travel. We can add a longevity term $\Lambda$ to our equation for the value of METI by defining $\Lambda$ as

$$\Lambda = \begin{cases} 0 & Lc < r_{nearest} \\ 1 & Lc \geq r_{nearest} \end{cases}, \quad (9)$$

where $c$ is the speed of light. When $Lc < r_{nearest}$, the watcher would be extinct by the time the reply arrived. For interstellar communication to occur between civilizations $Lc$ must be greater than $r_{nearest}$, at least for the particular civilizations involved. In the event of human extinction, any messages that had already been sent into space would be purely historical or archaeological artifacts. Although no new signals would be transmitted from Earth, these historical signals could potentially be detected by ETI and could therefore be valuable as a remnant of human civilization, contributing to the term $M_M$ regardless of whether or not a response is received. We will also define an additional analogous term $M_T$, which corresponds to the magnitude of the value of radio leakage into space that could potentially be observed after human extinction. This in turn allows us to redefine the expressions for the value of METI $U_M$ and the value of radio communication $U_R$:

$$U_M = p_d \langle M \rangle \Lambda + M_M - M_C, \quad (10)$$

$$U_R = p_d \langle M \rangle \Lambda + M_T + M_R, \quad (11)$$

This set of equations summarizes our framework for analyzing the benefits and harms from transmission into space.

## 5. Discussion

The framework established above describes the potential value of unintentional and deliberate transmissions into space that could be detected by ETI watchers. Although parameters such as the distribution of intelligent civilizations or the detection of a signal are highly uncertain, we can use this framework as a way to raise some critical questions regarding transmission into space.

Radio leakage has already given away the location of Earth in space to any nearby ETI watchers. If the transmission of signals from Earth is likely to bring about negative consequences because of detection by ETI (that is, if $\langle M \rangle < 0$), then it becomes important to decide whether or not radio communication should continue. It remains possible that SETI has failed to detect signals from ETI because most civilizations become radio quiet [4,31], perhaps because no one wants to "shout in the jungle" for too long [16]. This leads to our first question:

> *(Q1) Should human civilization cease radio communication on Earth in order to reduce the probability of contact with extraterrestrial civilizations?*

As stated above, because we cannot estimate the probability or magnitude of contact with ETI, we make no attempt to calculate the term $p_d \langle M \rangle$. By extension, any conclusions that depend on knowing $p_d \langle M \rangle$ are conditional.

In spite of the inherent uncertainty regarding the nature of ETI in the galaxy, we can proceed by assuming a working hypothesis for the valuation of radio communication on Earth today. One possible, and perhaps frequent, response to *(Q1)* is that the existence of terrestrial radio transmissions reduces more risk from other factors than increases the risk from ETI [5,20]. For example, military communication helps to maintain national security, and astronomical radar surveys help to identify potentially destructive impactors. We use this assumption as a working hypothesis *(H1)* that can be expressed as

$$(H1): M_R \gg |p_d \langle M \rangle \Lambda + M_T|, \quad (12)$$

which reduces Eq. (11) to $U_R \approx M_R$. Lack of knowledge regarding ETI in the galaxy prevents us from assessing whether or not *(H1)* is accurate; however in the absence of this knowledge it may be prudent for human civilization to adopt *(H1)* rather than cease radio communication altogether.[9] Indeed, humanity will not abandon radio transmissions given uncertain risks associated with ETI. Humanity may abandon radio communication in the future to favor some better, as-yet-unknown, technology. Earth would then enter a radio quiet phase. But until then, radio communication on Earth today is too valuable for humanity to be concerned with any potential harm from leakage into space.

If we assume that terrestrial radio operations are not be abandoned, then how should METI transmissions be assessed? Consider METI signals with a detectable volume less than that of radio leakage, such as high bandwidth broadcasts from transmitters at Arecibo, Goldstone, Evpatoria, or Madley 49 (Table 1). In this case, the probability of detection $p_d$ is less for METI signals than for radio leakage, so such METI attempts are probably unnecessary in order to attract attention from ETI. Such activities may still continue, perhaps because of other benefits from METI described by $M_M$, but they do not increase the risk described by the factor $p_d \langle M \rangle \Lambda$ in Eqs. (10) and (11). METI messages could include content more likely to be understood by ETI watchers compared with the leakage alone. However, as noted previously, even if ETI cannot understand a terrestrial transmission, the message would still reveal the presence of an intelligent civilization on Earth.

---

[9] As an illustrative example, consider the probability of correctly guessing the outcome of a coin toss. Under most circumstances we regard "heads" and "tails" as equally likely and disregard any other outcomes, such as the probability that a lion will eat us before the coin toss is complete. While such an unlikely scenario is possible, it may be a reasonable assumption to disregard this possibility.

Conducting METI at low levels of detectability (compared to radio leakage) also provides the advantage of low costs $M_C$, which allows for other benefits of METI $M_M$ such as engaging the public in constructing encoded messages for transmission to ETI [7] and conducting basic science related to METI in preparation for more powerful transmissions in the future. In this case, the value of transmission can be described according to Eq. (10) as $U_M \approx M_M$. If radio communication should continue according to (*H1*), then METI at low levels of detectability should also continue.

Different considerations apply to METI transmissions with a greater detectable volume than the leakage. For example, continuous low bandwidth broadcasts from Arecibo or Evpatoria would have a detectable volume $V$ significantly greater than television carrier waves (Table 1), at least in some parts of the sky. In this case, $p_d$ is greater for METI than for leakage. This leads to our second question:

*(Q2) Should human civilization transmit signals into space in attempt to initiate contact with extraterrestrial civilizations?*

We again invoke our working hypothesis (*H1*) for the value of radio communication on Earth such that $U_R \approx M_R$. This can be compared with the value of METI transmissions, for which we assume $U_M \approx p_d \langle M \rangle \Lambda$. The consequences of contact are thus the predominant term. A simple interpretation of these reduced equations suggests that METI at high levels of detectability should continue if $U_M > 0$ and cease if $U_M < 0$.

Given the wide range of possible outcomes for contact [26], it is premature to conclude that METI transmissions at a high level of detectability are necessarily a threat to human civilization. Answers to *(Q2)* strongly depend on personal expectations and beliefs about extraterrestrial civilizations, so the decision to transmit METI signals into space may span the boundaries of science, ethics, humanities, religion, and other relevant disciplines. Further exploration of the galaxy may also provide insight to *(Q2)*. For example, future missions such as the Terrestrial Planet Finder or New Worlds Observer may be able to observe the spectral signatures of extrasolar terrestrial planets to look for signs of biological life [65] or even technological ETI [66]. Ongoing SETI efforts also continue to search nearby stellar systems for communicative signals [67], while theoretical research may help to constrain the distribution of intelligent life in the galaxy [30,68-71] and could someday even help to characterize the nature of nearby ETI civilizations, if they exist. Such passive activities may also provide a way to better understand the distribution of ETI in the galaxy before beginning METI at high levels of detectability.

It is also possible that human civilization could become extinct for reasons other than contact with ETI. Global catastrophic risks include pandemics, climate change, global nuclear warfare, and near-Earth object impacts[10] [72], all of which could threaten the longevity of human civilization. Such threats may be a common obstacle to overcome for intelligent civilizations, including ours [73]. As described above, a situation where most intelligent civilizations in the galaxy fail to overcome existential threats is consistent with a small value of *L* in the Drake Equation (4). A value of *L* (multiplied by the speed of light) smaller than the distance $r_{nearest}$ to the nearest extraterrestrial civilization means that any transmissions, deliberate or unintentional, will never result in conversation with ETI (that is, $A = 0$). In this scenario, METI transmissions are futile because human civilization will become extinct before receiving a response. This situation corresponds to $p_d = 0$, so that Eqs. (10) and (11) simplify to $U_M = M_M - M_C$ and $U_R = M_T + M_R$, respectively. Radio communication still provides value $M_R$ to human civilization in this scenario, as do $M_T$ and $M_M$. Any transmission into space will persist after human extinction, so METI transmissions (as well as radio leakage) are analogous to a time capsule that provides information about a civilization that once existed. Even if human civilization only exists for a short time, it may still be worthwhile to transmit information into space as a way of preserving human cultural achievements. This is our next question:

*(Q3) Should human civilization transmit long-duration signals into space to preserve knowledge and history?*

It has been suggested that the scientific, technical, and cultural information of human civilization should be archived and protected in order to allow humanity to recover from a catastrophe [75,76]. Because transmissions into space propagate away from Earth at the speed of light, they cannot convey any information to future humans. However, if information regarding human civilization is received and comprehended by ETI, then this could count as positive value to the terms $M_M$ and $M_T$ because the preservation of human knowledge may eventually benefit humans that once existed but have now become extinct. In this case, the ability of ETI to decipher the information in a signal becomes archaeologically important, and deliberate METI broadcasts that use a carefully constructed language [55-58] will be more useful than radio leakage. This suggests that $M_M > M_T$, which implies a positive value for METI (at least if $M_C < M_M$) in a scenario with a low value of *L*. Although *(Q3)* may seem less pressing than other aspects of METI, it is worth considering transmissions into space as an archaeological artifact of human civilization.

Long-term METI can be a costly endeavor. A METI program that targets stars within 50 light years would need to transmit for at least 100 years before communication with ETI could occur, which would cost several million USD using a dedicated transmitter with a low (~0.1 Hz) bandwidth [2,22]. The costs of METI $M_C$ at high levels of detectability may not be small compared to the potential benefits of METI $p_d \langle M \rangle A + M_M$. Because of the uncertainty in the results of contact with ETI, it is difficult to estimate whether or not METI is worth the investment.

The merits of investing in METI are additionally difficult to evaluate because of the contrast between pessimistic scenarios that involve the extinction of humanity as a result of contact with ETI and optimistic scenarios involving contact that will bring about solutions to major problems of our civilization.[11] Furthermore, the costs of long-term METI at high levels of detectability likely would be much greater than the tangential benefits of METI (i.e. $M_C \gg M_M$), so the most relevant comparison is between $M_C$ the potential benefits or harms $p_d \langle M \rangle A$ from transmission. If contact with ETI never occurs, then the costs of METI programs might be considered wasteful. Alternatively, if the benefits of achieving remote contact with ETI roughly equal the costs spent in pursuit (i.e. $p_d \langle M \rangle \approx M_C$ so that $U_M \approx 0$), then the costs of METI can be compared to certain beneficial aspects of contact, such as new insights or technical abilities. However, it is difficult to assign cost values to any of these consequences. Perhaps the only benefit from METI will be the knowledge that ETI exist. Such knowledge would have tremendous value to scientists and academics and would also have profound implications for human belief systems [59-61], but not everyone may judge the philosophical implications of ETI worth a large investment in METI. Investment in large-scale METI represents a gamble with high costs and unknown rewards.

---

[10]Ongoing surveys using the Arecibo and Goldstone radars have concluded that there is little risk of civilization-ending asteroid impact for the next several hundred to a thousand years. Current efforts focus on identifying potential impactors that would cause only regional destruction, but these events would not contribute to human extinction [74].

## 6. Regulating Transmitting

If transmissions from Earth into space could potentially generate attention by extraterrestrial watchers, humanity should consider regulations to govern such transmissions. Existing laws and treaties do not adequately cover guidelines for transmission into space or the potential benefits or harms that could arise from such activities.

Since the beginning of modern SETI in 1959, there have been several proposals for an internationally-regulated framework to guide how detection would be handled. Recent efforts by the International Academy of Astronautics (IAA), the Institute for Space Law (IISL), the UN Committee on the Peaceful Use of Outer Space (COPUOS), the International Astronomical Union and others led to the "Declaration of Principles Concerning Activities Following the Detection of Extraterrestrial Intelligence" [78]. The document urges the utmost scientific rigor in examining any potential detection prior to making the announcement public, and recommends international consultation prior to transmitting any response. A draft declaration on "Principles Concerning the Sending of Replies to Extraterrestrial Intelligence" considers METI, but again concentrates on guidelines *following* detection.

There are no guidelines for METI specifically *prior* to detection. METI attempts to date have occurred without significant international consultation, using different encoding schemes, some of which are not likely to be intelligible to any watchers. It would be impossible to enforce a ban on any and all transmissions into space, but any large-scale efforts at METI would be responsible for how they represent humanity to any watchers. For this reason, Zaitsev [5,79] and Atri et al. [7] have proposed the development of a METI protocol to facilitate transmission strategy and standardize message composition. In practice, the decision to engage in large-scale METI may rest upon wealthy individuals or corporations with the motivation to pay for the power and technology.

---

[11]Strictly speaking, the possibility of beneficial contact requires only passive SETI observations if ETI are actively transmitting; however, ETI may not be transmitting any messages at all even if they are listening [77]. It is conceivable that ETI refrain from any interstellar messaging toward Earth until they receive a message from us, in which case METI would be required to achieve remote contact.

## 7. Conclusion

Both radio leakage and deliberate METI messages could be detected by any extraterrestrial watchers with sufficiently sensitive radio telescopes. Even if a signal cannot be interpreted, it provides evidence of a technological civilization on Earth.

Quantitatively assessing whether large-scale METI should continue in light of current radio communication strategies on Earth (and corresponding radio leakage) is hampered by two highly uncertain parameters: the probability of ETI detecting a transmission and the magnitude of response by ETI. Estimating these parameters is an exercise in speculation that we did not undertake. We posit that the benefits of radio communication on Earth today outweigh any benefits or harms that could arise from contact with ETI.

Transmissions from Earth that do not increase the probability of contact with ETI include METI transmissions that are short in duration, low in power or gain, or high in bandwidth because they are swamped by Earth's radio leakage. These transmissions create benefits such as opportunities for educational public outreach and the ability to develop scientific groundwork for future METI projects. The costs associated with METI at low levels of detectability are small, so such projects create overall positive value for humanity and should continue.

METI transmissions with a greater detectable volume than the radio leakage have a greater probability of being detected by extraterrestrial watchers, with highly uncertain outcomes. This uncertainty regarding contact with ETI creates difficulty in determining whether or not such METI attempts should occur. Existing governing structures are currently lacking for METI, so further thinking and discussion about METI is important. Even if we never succeed in receiving a message from an extraterrestrial civilization, METI may still prove a worthwhile investment as a way to increase humanity's awareness of itself in the greater cosmos.

**Acknowledgements**

The authors thank Milan Ćirković, Edward Schwieterman Jr., Megan Smith, and an anonymous reviewer for helpful suggestions. Any errors or opinions belong to the authors alone.

**Role of the Funding Source**

Lone Signal LLC provided funding for this research and requested that the Madley 49 transmitter be included in the analysis. The authors acknowledge scientific independence during the duration of this sponsored work. The design, methods, interpretation, writing, and submission of this study are the work of the authors alone and do not reflect, infer, or otherwise suggest any position held by Lone Signal LLC.